# Low-Power Cross-Phase Modulation in a Metastable Xenon-Filled Cavity for Quantum Information Applications

G. T. Hickman, T. B. Pittman, and J. D. Franson
*Department of Physics, University of Maryland Baltimore County,*
*Baltimore, MD 21250, USA*

Weak single-photon nonlinearities have many potential applications in quantum computing and quantum information. Here we demonstrate a relatively simple system for producing low-power cross-phase modulation using metastable xenon inside a high finesse cavity. The use of a noble gas such as xenon eliminates the contamination of the high-finesse mirrors that can occur when using alkali metal vapors such as rubidium. Cross-phase shifts of 5 mrad with 4.5 fJ control pulses were demonstrated. Numerical solutions of the master equation are in good agreement with the experimental results, and they predict that cross-phase shifts greater than 1 mrad per control photon should be achievable by reducing the size of the cavity.

## I. INTRODUCTION

Single-photon cross-phase shifts could be used to implement many operations that are needed for optical quantum communication and quantum computation [1-6]. Cross-phase shifts on the order of $\pi$ can be achieved using trapped atoms cooled to low temperatures. [7-10]. Although experiments of that kind have been very successful, they are relatively complex. Simpler and more robust ways to produce single-photon cross-phase shifts would be desirable for many practical applications outside of a controlled laboratory environment, such as quantum repeaters. Here we describe an approach that uses hot metastable Xe atoms in a high-finesse cavity to produce a cross-phase shift of 5 mrad with a 4.5 fJ control pulse. Weak cross-phase shifts of this magnitude can also be used for many quantum information applications [5-6, 11].

Weak cross-phase shifts have recently been generated using room-temperature rubidium vapor inside a hollow-core photonic bandgap fiber [12]. The use of a high-finesse cavity would be desirable, however, both to take advantage of the potential for further enhancement of the interaction strength and to avoid difficulties associated with the use of freely propagating beams [13]. A number of previous studies have investigated gas-filled Fabry-Perot cavities for low-power nonlinear optics, but deposition of the atomic medium onto the mirror surfaces has limited the attainable finesse [14-15]. The use of a noble gas such as xenon eliminates this difficulty.

We previously demonstrated nonlinear saturated absorption at low power levels using metastable Xe in a resonant cavity [16]. The 4.5 fJ control pulses used in this experiment correspond to approximately 18,000 photons inside the cavity. With several relatively simple improvements described in Section VI, this approach should be able to produce single-photon cross-phase shifts greater than 1 mrad, which would be large enough to be useful for applications in quantum communication and quantum computation [5-6].

The format of the remainder of this paper is as follows: In Section II we discuss the relevant properties of our high-finesse cavity and the transitions of interest in metastable Xe. Section III describes the experimental approach while Section IV presents a theoretical model that was used to calculate the expected cross-phase modulation. The experimental and theoretical results are compared in Section V and found to be in good agreement. Potential improvements to the approach are discussed in Section VI and a summary and conclusions are given in Section VII.

## II. METASTABLE XENON AND HIGH-FINESSE CAVITY

The lowest energy transition from the ground state of xenon is in the far ultraviolet and is not suitable for our cross-phase modulation experiments. Instead, we used a radio-frequency (RF) discharge to populate the $6s[3/2]_2$ Xe metastable state, which has an intrinsic lifetime of approximately 43 seconds and functioned as an effective ground state in our experiment [17]. As illustrated in Fig. 1, a pair of transitions are available from the metastable state in a ladder-type configuration. We chose to use the $6p[3/2]_2$ transition at 823 nm followed by the $8s[3/2]_1$ at 853 nm. For convenience we will designate these three states as $|0>$, $|1>$, and $|2>$, respectively.

A control light field tuned to the $|0>$ to $|1>$ transition can be used to produce a cross-phase shift on a probe (signal) tuned near the transition from $|1>$ to $|2>$. Using the upper transition for the signal has the advantage of producing very low loss in the absence of any control power. The transition dipole moments $\mu_{10}$ and $\mu_{21}$ were calculated using the lifetimes and branching ratios of the corresponding transitions [18]. For the hyperfine components used here this results in $\mu_{10} \approx 7.6 \times 10^{-30}$ C·m and $\mu_{21} \approx 1.2 \times 10^{-30}$ C·m [19-21]. The available branching ratios for the upper transition were relatively uncertain and the estimated dipole moments are based in part on two-photon absorption measurements performed in our laboratory. These dipole moments are roughly comparable to those of the commonly used transitions in Rb, with $\mu_{21}$ being somewhat smaller.

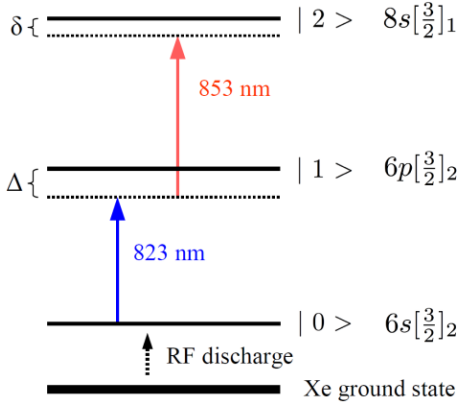

Fig. 1. Xenon energy level diagram showing the levels and transitions used in our experiment. The dipole matrix elements for the first and second transitions are $\mu_{10} \approx 7.6 \times 10^{-30}$ C·m and $\mu_{21} \approx 1.2 \times 10^{-30}$ C·m, respectively. The parameters $\Delta$ and $\delta$ represent the frequency detunings from states $|1\rangle$ and $|2\rangle$, respectively.

A pair of super-polished dielectric mirrors was mounted inside a vacuum chamber filled with 1 Torr of Xe gas. The mirrors formed a confocal cavity with a finesse of approximately 3,000, a length of 25 mm, and a beam waist radius of 60 μm. The measured quality factor was $Q = 3 \times 10^8$. The resonant frequency of the cavity was tuned by varying the temperature of the mounting fixture as described in more detail in Ref. [16]. The RF discharge used to excite the Xe atoms into the metastable state produced no noticeable degradation of the cavity finesse.

### III. EXPERIMENTAL DESIGN

Fig. 2 shows an overview of the experiment design. Two tunable diode lasers (Toptica DL pro) were tuned to 823 nm and 853 nm to produce the control and probe beams, respectively. Each laser passed through a set of amplitude modulators capable of producing pulses of 30 to 60 ns duration. A pair of photodetectors labeled D1 and D2 monitored the two beams to ensure proper biasing of the amplitude modulators. The frequencies of both beams were continuously monitored using a high-precision wavelength meter (HighFinesse WSU30) with a calibrated accuracy of 30 MHz.

To facilitate high speed locking of the two laser beams to the desired detunings, two high bandwidth photodetectors measured the transmission of the beams through the cavity (for reasons of clarity these detectors are not shown in Fig. 2). Relatively high intensities of the two beams were required in order to produce a sufficiently large signal at the detectors. To accomplish this, the control beam was divided into two separate paths using a set of fiber-coupled optical switches (Thorlabs OSW12-830E) that controlled which path the beam would take. A variable attenuator was added to one of the paths for the low-intensity measurements, while the higher intensity in the other path was used to periodically lock the laser frequency to the desired detuning.

The cross-phase shift in the signal beam was measured using the homodyne detection technique shown in the right-hand side of Fig. 2, where the signal interferes with a much stronger local oscillator beam in order to reduce the effects of detector noise. The weak signal and the strong local oscillator beam propagated in opposite directions through a Sagnac loop interferometer and interference between them was measured in the two output ports using balanced photodetectors D3(a) and D3(b). A Sagnac interferometer was used due to its high intrinsic phase stability. The control pulses were timed to reach the cavity at the same time as the clockwise-propagating 853 nm probe pulses to produce a cross-phase shift, while the counterclockwise-propagating local oscillator pulses passed through the cavity several hundred nanoseconds later without being phase-shifted. The Sagnac loop was implemented using 150 m of polarization-preserving optical fiber.

An isolator inside the Sagnac loop attenuated the clockwise-propagating 853 nm probe pulses to an intensity that was sufficiently weak for them to interact with the control pulses in the cavity. The counter-clockwise propagating 853 pulses were not attenuated by the isolator, which allowed them to function as a strong local oscillator. A time-dependent phase modulator was included in the loop and used to impart a 90° shift on one but not both of the counter-propagating pulses, which maximized the sensitivity of the output interference pattern to any additional small relative phase shifts.

The presence of a large number of fiber-coupled optical components in the beam path and the use of short pulses to excite the high-finesse cavity resulted in large optical losses. To counteract these losses, a pair of tapered amplifiers (Thorlabs TPA850P10) was used to increase the power of the 853 nm beam as required for the local oscillator. A Pockels cell was placed after the amplifiers to prevent amplified spontaneous emission from interfering with the measurement.

A balanced photodetector (Thorlabs PDB420A) was used to measure the interference between the probe and local oscillator. This signal was proportional to the nonlinear cross-phase shift of interest. Because the pulses used to excite the cavity were shorter than the cavity lifetime of 80 ns, the fraction of incident light coupled into the cavity was relatively small and the majority of each incoming pulse was reflected from the cavity surface. To minimize the effect of these reflections on the signal, the geometry of the Sagnac loop was designed in such a way that the back-reflected pulses arrived at the balanced detector D3 at a different time than the signal and local oscillator pulses. This allowed a fast balanced photoreceiver (75 MHz bandwidth) to be used in combination with a nanosecond analog-to-digital converter (FAST ComTec 7072) to sample the cross-phase shift signal while ignoring the reflected pulses.

The data collection system operated at a repetition rate of 200 kHz using high-speed NIM-bin electronics. In order to further reduce the effects of low-frequency amplifier noise and back-reflections, each signal pulse was followed immediately (within a few microseconds) by a second signal pulse but with the control pulse turned off. The results from these two measurements were subtracted to reduce any spurious effects. An average over approximately $10^5$ such measurements was used to estimate the cross-phase shift due to the presence of the control beam.

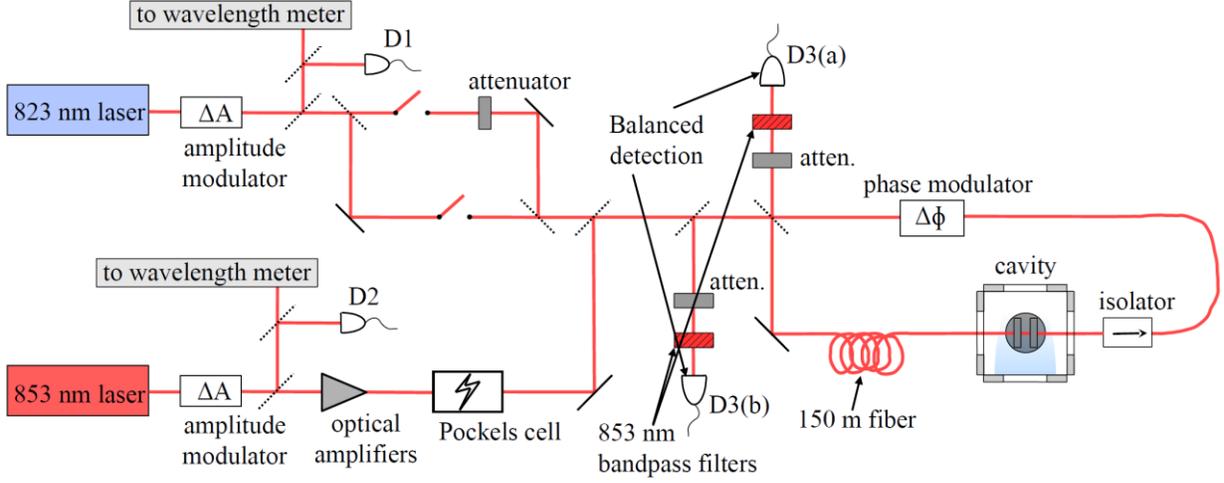

Fig. 2. Overview of the experimental design of the cross-phase shift measurements. The measured phase shift from the homodyne detector was proportional to the interference between counter-propagating signal and local oscillator pulses inside a Sagnac interferometer containing the high-finesse cavity. Control pulses at 823 nm were timed to produce a cross-phase shift on only one of the two counter-propagating 853 nm pulses inside the Sagnac loop. The other 853 nm pulse served as the local oscillator and the relative phase shift was measured using balanced detectors (D3) at the output ports of the Sagnac interferometer.

The relative timing of the pulses and analog-to-digital acquisition windows had to be carefully adjusted. Fig. 3 shows an oscilloscope trace of the output of the balanced detector D3 during a calibration run. For this test, the phase modulator within the Sagnac loop was used to simulate the effects of a cross-phase shift by applying an extra 90° phase shift to the clockwise-propagating probe pulse, with the control beam turned off. After each cycle the measurement test was repeated with no extra shift applied, thus simulating the effects of an actual cross-phase shift measurement. The results of this procedure were used to calibrate the sensitivity of the phase shifts as measured by the difference between the D3 output voltages for the two cases. Fig. 3 also illustrates the relative timing of the cross-phase modulation signal and the analog-to-digital acquisition time.

## IV. THEORETICAL MODEL

The expected cross-phase shift was calculated using a semi-classical density matrix calculation in which the optical pulses were treated as classical light fields while the Xe atoms were described by a 3-level open quantum system. This approach is valid for the photon numbers used in this experiment, while similar results should be expected in the single-photon regime with the cross-phase shift proportional to the control beam intensity.

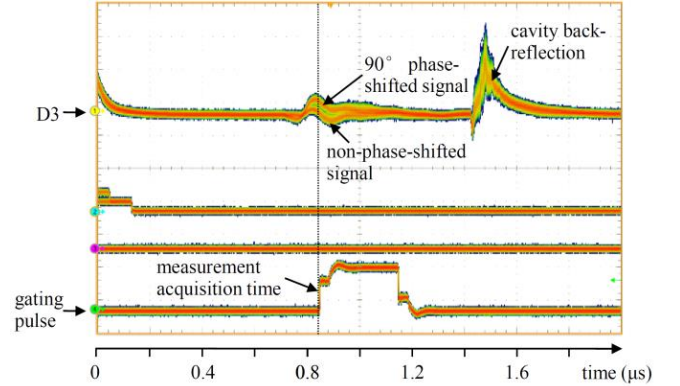

Fig. 3. Typical oscilloscope trace showing the relative timing of the measured phase shift and the analog-to-digital acquisition time during a calibration run. The topmost trace shows the signal as seen by the balanced photodetector (D3 in Fig. 2), while the bottom trace shows the gating pulse used to set the time at which the high-speed analog-to-digital converter acquired the measured voltage. In this test an additional 90° phase shift was alternately applied and then not applied to the 853 nm pulses. The difference between the two resulting traces produced a visible phase shift measurement signal.

Because the two-photon interaction took place in a standing-wave cavity, it consisted of both a counter-propagating Doppler-free part and a co-propagating Doppler-broadened part. The Doppler-free contribution dominates for small detunings near a two-photon resonance and our analysis neglected the much-smaller Doppler-broadened contribution. For simplicity the left- and right-travelling probe beams were also assumed to interact with independent atomic ensembles. Using basis states that rotate along with the driving fields, the resulting master equation for a Xe atom inside the cavity mode can be written as [22]:

$$\begin{aligned}
\dot{\sigma}_{00} &= \Gamma_{10}\sigma_{11} - \left(iR_{10}\sigma_{10}^{*} - iR_{10}^{*}\sigma_{10}\right) \\
\dot{\sigma}_{11} &= -\gamma_{1}\sigma_{11} + \Gamma_{21}\sigma_{22} + \left(iR_{10}\sigma_{10}^{*} - iR_{10}^{*}\sigma_{10}\right) - \left(iR_{21}\sigma_{21}^{*} - iR_{21}^{*}\sigma_{21}\right) \\
\dot{\sigma}_{22} &= -\gamma_{2}\sigma_{22} + \left(iR_{21}\sigma_{21}^{*} - iR_{21}^{*}\sigma_{21}\right) \\
\dot{\sigma}_{10} &= -\gamma_{10}\sigma_{10} + i\Delta\sigma_{10} - iR_{10}\sigma_{11} + iR_{10}\sigma_{00} + iR_{21}^{*}\sigma_{20} \\
\dot{\sigma}_{21} &= -\gamma_{21}\sigma_{21} + i(\delta - \Delta)\sigma_{21} - iR_{21}\sigma_{22} + iR_{21}\sigma_{11} - iR_{10}^{*}\sigma_{20} \\
\dot{\sigma}_{20} &= -\gamma_{20}\sigma_{20} + i\delta\sigma_{20} + iR_{21}\sigma_{10} - iR_{10}\sigma_{21}
\end{aligned} \quad (1)$$

Here the $\sigma_{ji}$ are the density matrix elements in the rotating basis, $\gamma_{i}$ is the inverse of the state $|i\rangle$ lifetime, and $\gamma_{ji} = (\gamma_{j} + \gamma_{i})/2$ are the dephasing rates for the off-diagonal elements of $\hat{\sigma}$. The broadening due to atomic collisions and to the presence of the RF discharge field was small compared with the natural linewidths of the transitions and was neglected. $\Gamma_{ji}$ is the $|i\rangle$ to $|j\rangle$ spontaneous transition rate while $R_{10} = \mu_{10}E_{c}(t)/\hbar$ and $R_{21} = \mu_{21}E_{p}(t)/\hbar$ are the electromagnetic coupling strengths for the two transitions in the presence of the laser fields. Here $E_{c}$ and $E_{p}$ designate the complex electric field amplitude of the control and probe beams, respectively [22].

The parameters $\Delta$ and $\delta$ are the detunings in rad/s from states $|1\rangle$ and $|2\rangle$. The field amplitude was approximated by a constant value across an effective cavity mode volume, as described in Ref. [23]. The decay and transition rates were calculated using two-photon absorption measurements performed in our lab, combined with published data for the state lifetimes and branching ratios [19-21]. The resulting values were $\gamma_{1} = 32$ MHz, $\gamma_{2} = 14$ MHz, $\Gamma_{10} = 29$ MHz, $\Gamma_{21} = 65$ kHz, $\mu_{10} = 7.6 \times 10^{-30}$ C·m and $\mu_{21} = 1.2 \times 10^{-30}$.

The electric fields and electromagnetic coupling strengths in Eq. (1) were time-dependent due to the use of pulsed signal and control beams. In the limit of a small cavity with high finesse, the time dependence of the field amplitudes is given by

$$\begin{aligned}
\dot{E}_{c} &= \left(t_{c}E_{c}^{0}(t) + \left(r_{c}^{2}e^{(-\beta_{c}+i\phi_{c})\tau_{c}} - 1\right)E_{c}\right)/\tau_{c} \\
\dot{E}_{p} &= \left(t_{p}E_{p}^{0}(t) + \left(r_{p}^{2}e^{(-\beta_{p}+i\phi_{p})\tau_{p}} - 1\right)E_{p}\right)/\tau_{p}.
\end{aligned} \quad (2)$$

Here $E_{i}^{0}(t)$ represents the electric field amplitudes of the control and probe input pulses that are incident on the cavity while $t_{i}$ and $r_{i}$ are the mirror reflection and transmission coefficients, respectively, which are assumed to be the same for both mirrors. From the observed value of the quality factor Q, $t_{i} = 0.9995$ and $r_{i} = 0.0316$ for both wavelengths. The parameter $\tau_{i}$ is the time required for one round trip of field $i$ within the cavity.

Equation (2) can be derived by considering the changes in the field during a single round trip through the cavity. The constants $\beta_{i}$ and $\phi_{i}$ designate the field decay rates and phase shifts due to the interaction with the Xe atoms and are given by

$$\begin{aligned}
\phi_{i} &= \omega_{i} \cdot \mathrm{Re}\{\chi_{i}\}/2 \\
\beta_{i} &= \omega_{i} \cdot \mathrm{Im}\{\chi_{i}\}/2
\end{aligned} \quad (3)$$

where $\omega_{i}$ is the angular frequency of beam $i$ and $\chi_{i}$ is its susceptibility. Rather than calculating $\chi_{i}$ for each atom separately, we first considered the case of a single atom and then multiplied the results by the effective number of interacting atoms. This approach is valid provided that the density of atoms is sufficiently small, which was the case in our experiment.

Eq. (1) is then coupled to Eq. (2) through the values of $\beta_{i}$ and $\phi_{i}$, where [22]

$$\begin{aligned}
\chi_{c} &= N \cdot \frac{\sigma_{10}\mu_{10}^{*}}{E_{c}} \\
\chi_{p} &= N \cdot \frac{\sigma_{21}\mu_{21}^{*}}{E_{p}}.
\end{aligned} \quad (4)$$

Here N is the density of metastable Xe atoms. Cross-phase modulation of the control pulses due to the presence of the probe was negligible and was ignored. This corresponds to using $\phi_{c} = 0$ in the theoretical model.

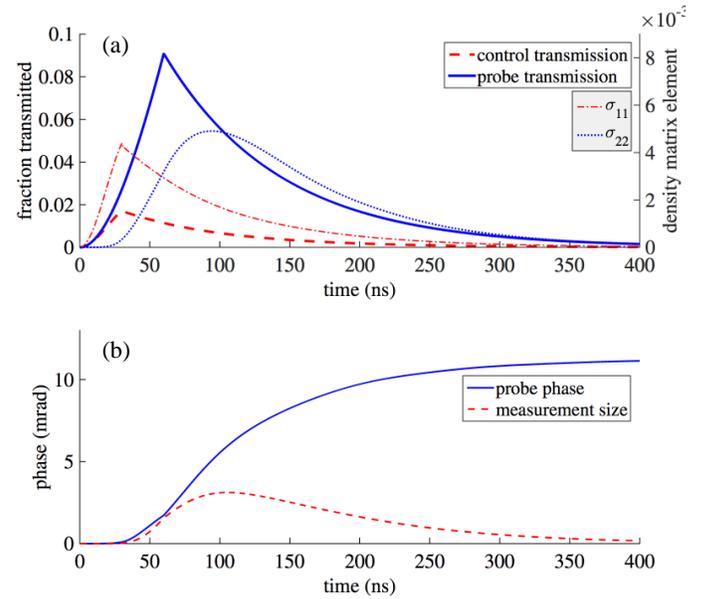

Fig. 4. Results of the density matrix calculation for a typical set of parameters. (a) The occupation probabilities of the excited atomic states $|1\rangle$ and $|2\rangle$ plotted along with the fraction of the incident power transmitted through the cavity for the control and probe pulses. (b) The cross-phase shift acquired by the probe pulse (solid line) and the product of the cross-phase shift multiplied by the intensity of the probe pulse as it leaves the cavity (dashed line), which is proportional to the output of the balanced detector. The scaling for the y-axis of the dashed curve in (b) is arbitrary.

The Doppler broadening of the atomic linewidths was included in the calculations using a Monte Carlo method in which a random set of atomic velocity groups was sampled from a Gaussian distribution. The distribution width was determined from the measured Doppler width of the 823 nm transition, which was 440 MHz full-width-at-half-maximum (FWHM). Eqs. (1) through (4) were solved numerically for each velocity group and an average was taken over the Doppler-broadened ensemble.

## V. COMPARISON OF THEORETICAL AND EXPERIMENTAL RESULTS

The results from the density matrix calculation for a typical set of parameters are shown in Fig. 4. The detuning $\Delta$ of the control beam was chosen to be $\Delta = -800$ MHz while the two-photon detuning $\delta$ was varied to maximize the induced cross-phase shift, as was done in the experimental measurements as well. The durations of the probe and control pulses were chosen to be 60 ns and 30 ns, respectively, which were the values used in the experiment. The atomic decay and transition rates, dipole moments, and measurement acquisition time used in the calculations were also the same as in the experiment. The effective density of metastable Xe atoms and the delay time between the control and probe pulses were varied within the experimental uncertainties to give the best fit with the measured data.

In addition to the output field amplitudes and atomic populations shown in Fig. (4a), the calculated cross-phase shift (solid line) is shown as a function of time in Fig. (4b). The output signal from the homodyne detector corresponds to the product of the phase shift and the amplitude of the probe beam leaving the cavity, and is shown by the dashed line in Fig (4b). It can be seen that the magnitude of the calculated cross-phase shift gradually increases as a function of time, but that the dependence of the homodyne signal on the amplitude of the probe beam gives a maximum value of the homodyne signal at a measurement time of approximately t = 100 ns after the arrival of the incident pulses. Subsequent measurements have a lower signal-to-noise ratio even though the phase shift is larger. As a result, the signal acquisition times were chosen to be near 100 ns.

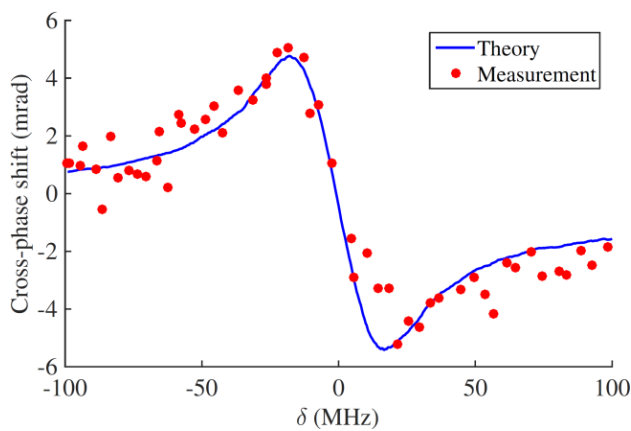

Fig. 5. Comparison of the measured cross-phase shift with the theoretical prediction from the density matrix calculation. A constant background was subtracted from the experimental data to remove a small bias produced by back-reflections of the control beam. The maximum phase shift observed was approximately 5 mrad.

Fig. 5 shows the results of the cross-phase shift measurements obtained under the conditions described above. A maximum cross-phase shift of approximately 5 mrad was observed using 4.5 fJ control pulses. The noise in the data is primarily due to electronic noise from the balanced detector. These results correspond to an average of approximately 18,000 control photons per pulse, or 0.3 µrad of cross-phase modulation per photon. It can be seen that the experimental and theoretical results are in reasonably good agreement.

More systematic measurements of the cross-phase shift as a function of other experiment parameters would be desirable. This was not possible using the current apparatus because the temperature control of the cavity length had a long time constant and could not compensate for short-term variations in the resonant frequency. In addition, the resonant frequency of the cavity was shifted by a small amount depending on the RF power level, which made it difficult to measure the effects of varying metastable xenon density. Both of these problems could be addressed by using piezoelectric control of the cavity length.

## VI. DISCUSSION AND POSSIBLE IMPROVEMENTS

Quantum computation and quantum communication protocols based on a weak Kerr nonlinearity typically require single-photon cross-phase shifts on the order of 1 mrad [5-6], which is several orders of magnitude larger than that demonstrated in this experiment. Here we discuss several potential improvements to the apparatus that would enable the system to produce nonlinear phase shifts of the required magnitude.

Single-photon nonlinearities in a high-finesse cavity are roughly proportional to $Q/V$, where $V$ is the effective mode volume. It would be relatively straightforward, for example, to decrease the cavity length by a factor of 10 to 2.5 mm, which would also decrease the mode diameter by a factor of $\sqrt{10}$ to give a factor of $10\sqrt{10} = 32$ decrease in the mode volume. Increasing the finesse of the mirrors by a factor of 10 to 30,000 at the same time would maintain the same value of Q. Thus it should be possible to substantially increase the single-photon cross-phase shift by reducing the mirror separation, with an expected enhancement of three orders of magnitude for a cavity length of 250 µm.

The strength of the upper atomic transition was found to be significantly smaller than we had expected. An inconsistency in the published transition rates and associated excited state lifetimes made it difficult to obtain accurate values for the dipole moments [20-21]. The square of the upper-transition dipole moment $|\mu_{21}|^2$, which is proportional to the expected cross-phase modulation, appears to be a factor of approximately 7 larger for the transition to the $|2\rangle = 8s[3/2]_2$ level at 862 nm than it is for the transition to the $8s[3/2]_1$ level at 853 nm in our current experiment. Thus an order of magnitude increase in the cross-phase shift should be achievable using a different set of transitions in metastable xenon. Preliminary results using this set have already shown an increase in the cross phase shift by a factor of two.

With the above-mentioned changes our system should be able to produce single-photon cross-phase shifts on the order of milliradians. If further improvement is desired then it may be necessary to modify our system to use a lambda-type transition. For example, the counter-propagating beams in our cavity are only approximately Doppler-free due to the 3% difference in the wavelengths of the control and probe beams. The effects of this residual Doppler width are illustrated in Fig. 6, which compares the calculated cross-phase shift with and without a residual Doppler shift of this magnitude. It can be seen that a factor of approximately three enhancement in the cross-phase shift could be obtained if the wavelengths of the control and probe beams were more nearly the same. Using a lambda-type transition between the hyperfine levels of metastable xenon-129 could accomplish this, as illustrated in Fig. 7. This approach also has the advantage that it does not depend on the relatively small value of $\mu_{21}$ for the 853 nm transition.

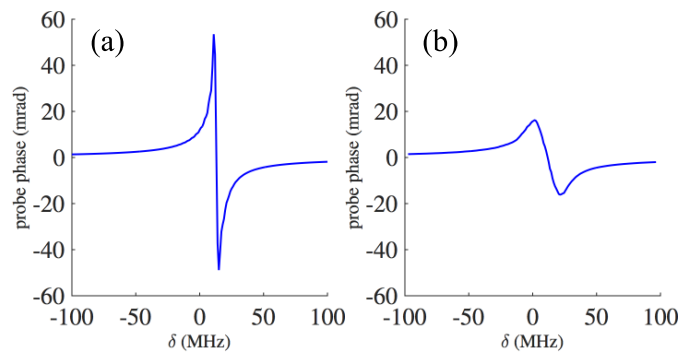

Fig. 6. Typical simulated phase shift measurements vs. detuning $\delta$ when the two-photon transition is taken to be (a) fully Doppler-free and (b) influenced by a 3% (16 MHz FWHM) residual Doppler broadening, assuming counter-propagating control and probe beams. The cross-phase modulation is large and sharply peaked for the Doppler-free case, while the Doppler-broadened spectrum is shallower. The phase shift values are much larger than those of Fig. 5 because the measurements here were taken 350 ns after the beginnings of the probe pulses, whereas the corresponding wait time used in Fig. 5 was 60 ns.

The use of a lambda transition has the disadvantage of relatively large loss for the probe beam unless the hyperfine levels can all be initially pumped into state $|0\rangle$. Simulations performed for this set of transitions using a cavity length of 2.5 mm with a finesse of 30,000 predict an achievable single-photon cross-phase shift of 0.6 mrad.

## VII. CONCLUSIONS

In summary, we have demonstrated a relatively simple technique for producing ultra-low power nonlinear cross-phase shifts using metastable Xe inside a high-finesse cavity. The use of a noble gas such as xenon eliminates the degradation of the high-finesse mirrors that often occurs when using alkali metals such as rubidium [24]. Phase shifts of 5 mrad were demonstrated using a control field with 4.5 fJ per pulse, which corresponds to approximately 18,000 photons inside the cavity. A numerical solution to the master equation for the xenon atoms inside the cavity was in good agreement with the experimental results.

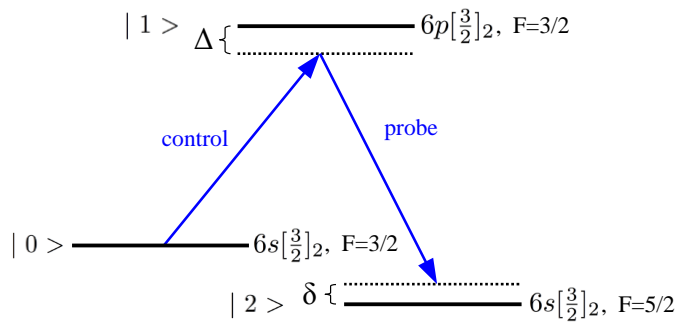

Fig. 7. Lambda-type energy level diagram for the production of a cross-phase shift using the hyperfine levels of metastable $^{129}$Xe. This approach is essentially Doppler-free and it takes advantage of the relatively large dipole matrix element for the 6s[3/2]$_2$ to 6p[3/2]$_2$ transition in Xe of $2.4 \times 10^{-29}$ C·m. The parameters $\Delta$ and $\delta$ again represent the detunings from states $|1\rangle$ and $|2\rangle$, respectively.

Our calculations show that it should be possible to produce much larger single-photon phase shifts by reducing the length of the cavity and by using a different ladder transition in xenon. Cross-phase shifts of that magnitude could be used to implement QKD and quantum logic operations. This approach would allow a relatively simple and rugged implementation that may be required for practical applications outside of the laboratory, such as quantum repeaters.

## ACKNOWLEDGEMENTS

The authors would like to acknowledge valuable discussions with B. T. Kirby, C. Broadbent, J. Howell, and S. Sergienko. This work was supported in part by DARPA DSO Grant No W31P4Q-12-1-0015 and by NSF grant No. PHY-1402708.